\pdfoutput=1
\RequirePackage{ifpdf}
\ifpdf 
\documentclass[pdftex]{sigma}
\else
\documentclass{sigma}
\fi

\numberwithin{equation}{section}

\newtheorem*{Theorem*}{Theorem}

 { \theoremstyle{definition}

 }

\usepackage{mathrsfs,amscd}
\usepackage[svgnames,table]{xcolor}
\usepackage{booktabs}

\usepackage{tikz,pgfplots}
\usetikzlibrary{calc,cd}

\newcommand{\Ggr}{\mathcal{G}}
\newcommand{\Hgr}{\mathcal{H}}
\newcommand{\Kgr}{\mathcal{K}}

\newcommand{\Bgr}{\mathcal{B}}
\newcommand{\g}{\mathfrak{g}}
\newcommand{\h}{\mathfrak{h}}

\renewcommand{\d}{\partial}

\newcommand{\so}{\mathfrak{so}}

\renewcommand{\k}{\mathfrak{k}}

\newcommand{\m}{\mathfrak{m}}

\newcommand{\x}{\boldsymbol{x}}
\newcommand{\X}{\boldsymbol{X}}
\newcommand{\y}{\boldsymbol{y}}

\newcommand{\w}{\boldsymbol{w}}
\ifdefined\C
\renewcommand{\C}{\boldsymbol{C}}
\else
\newcommand{\C}{\boldsymbol{C}}
\fi
\newcommand{\J}{\boldsymbol{J}}
\newcommand{\B}{\boldsymbol{B}}

\renewcommand{\P}{\boldsymbol{P}}

\newcommand{\eL}{\mathscr{L}}

\newcommand{\ad}{\operatorname{ad}}

\newcommand{\RR}{\mathbb{R}}

\newcommand{\MM}{\mathbb{M}}

\newcommand{\zdS}{\mathsf{dS}}
\newcommand{\zAdS}{\mathsf{AdS}}

\newcommand{\zG}{\mathsf{G}}

\newcommand{\zAdSG}{\mathsf{AdSG}}
\newcommand{\zdSG}{\mathsf{dSG}}
\newcommand{\ztAdSG}{\mathsf{AdSG}}
\newcommand{\ztdSG}{\mathsf{dSG}}
\newcommand{\pG}{\mathsf{PG}}
\newcommand{\pAdSG}{\mathsf{PAdSG}}
\newcommand{\pdSG}{\mathsf{PdSG}}
\newcommand{\ptAdSG}{\mathsf{PAdSG}}

\begin{document}

\newcommand{\arXivNumber}{2208.07627}

\renewcommand{\PaperNumber}{035}

\FirstPageHeading

\ShortArticleName{From pp-Waves to Galilean Spacetimes}

\ArticleName{From pp-Waves to Galilean Spacetimes}

\Author{Jos\'e FIGUEROA-O'FARRILL~$^{\rm a}$, Ross GRASSIE~$^{\rm b}$ and Stefan PROHAZKA~$^{\rm a}$}

\AuthorNameForHeading{J.~Figueroa-O'Farrill, R.~Grassie and S.~Prohazka}

\Address{$^{\rm a)}$~Maxwell Institute and School of Mathematics, The University of Edinburgh,\\
\hphantom{$^{\rm a)}$}~James Clerk Maxwell Building, Peter Guthrie Tait Road, Edinburgh EH9 3FD, Scotland, UK}
\EmailD{\href{mailto:j.m.figueroa@ed.ac.uk}{j.m.figueroa@ed.ac.uk}, \href{mailto:stefan.prohazka@ed.ac.uk}{stefan.prohazka@ed.ac.uk}}

\Address{$^{\rm b)}$~Laboratory for Foundations of Computer Science, School of Informatics,\\
\hphantom{$^{\rm b)}$}~The University of Edinburgh, Informatics Forum, 10 Crichton Street, Edinburgh EH8 9AB,\\
\hphantom{$^{\rm b)}$}~Scotland, UK}
\EmailD{\href{mailto:rgrassie@ed.ac.uk}{rgrassie@ed.ac.uk}}

\ArticleDates{Received October 27, 2022, in final form May 22, 2023; Published online June 03, 2023}

\Abstract{We exhibit all spatially isotropic homogeneous Galilean spacetimes of dimension $(n+1) \geq 4$, including the novel torsional ones, as null reductions of homogeneous pp-wave spacetimes. We also show that the pp-waves are sourced by pure radiation fields and analyse their global properties.}


\Classification{22F30; 53C30; 53Z05}

\section{Motivation and introduction}

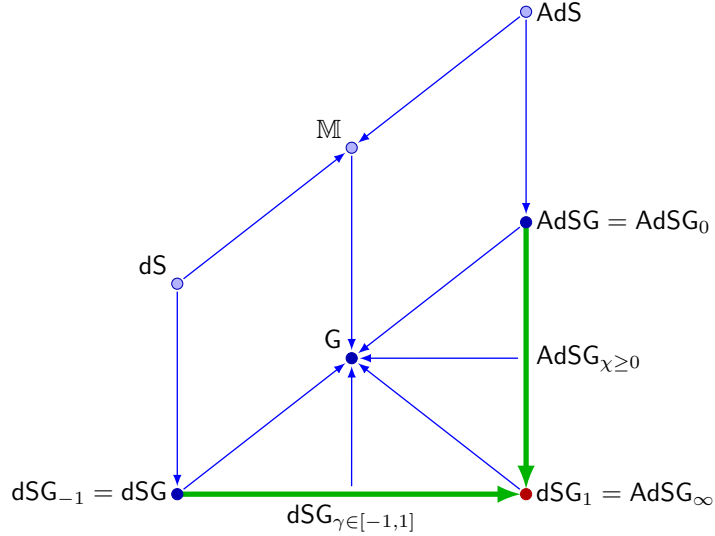
\begin{figure}[h!] \centering
 \begin{tikzpicture}[scale=1.4,>=latex, shorten >=3pt, shorten <=3pt, x=1.0cm,y=1.0cm]
 \coordinate [label=above left:{\small $\zdS$}] (ds) at (5.688048519056286,-0.5838170592960186);
 \coordinate [label=above left:{\small $\MM$}] (m) at (7.344024259528143,0.7080914703519907);
 \coordinate [label=right:{\small $\zAdS$}] (ads) at (9,2);
 \coordinate [label=left:{\small $\zdSG_{-1} = \zdSG$}] (dsg) at (5.688048519056286, -2.5838170592960186);
 \coordinate [label=right:{\small $\ztdSG_1 = \ztAdSG_\infty$}] (dsgone) at (9, -2.5838170592960186);
 \coordinate [label=above left:{\small $\zG$}] (g) at (7.344024259528143, -1.2919085296480093);
 \coordinate [label=right:{\small $\zAdSG=\ztAdSG_0$}] (adsg) at (9,0);
 %
 %
 \coordinate [label=below:{\small $\ztdSG_{\gamma\in[-1,1]}$}] (tdsg) at (7.344024259528143, -2.5838170592960186);
 \coordinate [label=right:{\small $\ztAdSG_{\chi\geq0}$}] (tadsg) at (9, -1.2919085296480093);
 %
 %
 \draw [->,line width=0.5pt,color=blue] (dsgone) -- (g);
 \draw [->,line width=0.5pt,color=blue] (7.344024259528143,-2.5838170592960186) -- (g);
 \draw [->,line width=0.5pt,color=blue] (9, -1.2919085296480093) -- (g);
 %
 %
 \draw [->,line width=0.5pt,color=blue] (adsg) -- (g);
 \draw [->,line width=0.5pt,color=blue] (dsg) -- (g);
 \draw [->,line width=0.5pt,color=blue] (ads) -- (m);
 \draw [->,line width=0.5pt,color=blue] (ds) -- (m);
 \draw [->,line width=0.5pt,color=blue] (ds) -- (dsg);
 \draw [->,line width=0.5pt,color=blue] (m) -- (g);
 \draw [->,line width=0.5pt,color=blue] (ads) -- (adsg);
 %
 \begin{scope}[>=latex, shorten >=0pt, shorten <=0pt, line width=2pt, color=green!70!black]
 \draw [->,shorten >=2pt] (adsg) --(dsgone);
 \draw [->,shorten >=2pt] (dsg) --(dsgone);
 \end{scope}
 \filldraw [color=red!70!black,fill=red!70!black] (dsgone) circle (1.5pt);
 \foreach \point in {g,adsg,dsg}
 \filldraw [color=blue!70!black,fill=blue!70!black] (\point) circle (1.5pt);
 \foreach \point in {m,ds,ads}
 \filldraw [color=blue!70!black,fill=blue!30!white] (\point) circle (1.5pt);
 \end{tikzpicture}
 \caption{Spatially isotropic homogeneous Galilean spacetimes in
 dimension $n+1 \geq 4$. The blue arrows denote limits arising from
 Lie algebra contractions and show that (A)dS--Galilei arise
 naturally as limits of (A)dS, while the families of torsional
 Galilean spacetimes (green arrows) do not.}
 \label{fig:Galilean}
\end{figure}

Maximally symmetric Lorentzian spacetimes provide a natural arena for
many areas of physics ranging from high energy physics (Minkowski
space) to cosmology (de Sitter space) and quantum gravity (anti-de
Sitter space). Beyond these well-known Lorentzian spacetimes, there is
a~plethora of equally interesting Galilean, Carrollian and
Aristotelian spacetimes. These spacetimes were classified in
\cite{Figueroa-OFarrill:2018ilb}, completing the pioneering work of
Bacry and L\'evy-Leblond \cite{Bacry:1968zf}. One of the salient
features of the classification in~\cite{Figueroa-OFarrill:2018ilb} is
the emergence of two novel families of generically
torsional\footnote{These reductive homogeneous spacetimes are
 torsional in the sense that their canonical invariant connections
 have nonzero torsion. This should not be confused with the
 \emph{intrinsic} torsion \cite{Figueroa-OFarrill:2020gpr, MR334831}
 of the Galilean geometry, which is given by the exterior derivative
 of the clock one-form. The intrinsic torsion vanishes in all cases
 considered, but that does not mean that every connection has
 vanishing torsion, just that there is at least one connection which
 does.} Galilean spacetimes, containing the (A)dS--Galilei spacetimes
as special (non-torsional) points, see Figure~\ref{fig:Galilean}.

Unlike many of the spacetimes in the classification, very little is
known about these torsional Galilean spacetimes; however, there are
various lines of investigation in which they may play interesting
roles.

{\it Holography.} Particles in (A)dS--Galilei spacetimes
have recently appeared as instructive toy models in the
holography literature~\cite{Bizon:2018frv,Maxfield:2022hkd,
 Susskind:2021omt}. It is, therefore, natural to ask how these
torsional geometries, which interpolate between AdS--Galilei and
dS--Galilei, could appear in these models.

{\it Intrinsically Galilean.} The torsional Galilei algebras are a
purely Galilean feature with no Lorentzian or Carrollian
counterpart~\cite{Figueroa-OFarrill:2018ilb}. They are the
only one-parameter family of non-equivalent maximally
symmetric spaces, and they do not arise naturally as limits
of Lorentzian spaces. Therefore, it could well be that
effects due exclusively to the presence of torsion may
be intrinsically Galilean, in some sense.

This paper and the accompanying work~\cite{Figueroa-OFarrill:2022tlf}
aim to deepen our understanding of these novel geometries and
demystify them.

It is well known that we can obtain non-relativistic spacetimes by
null reductions of Bargmann spacetimes in one dimension higher
\cite{PhysRevD.31.1841,Julia:1994bs}; namely, Lorentzian spacetimes
admitting a nowhere-vanishing null vector field. In Duval, Burdet,
K\"unzle and Perrin~\cite{PhysRevD.31.1841} this null vector field is
assumed to be parallel with respect to the Levi-Civita connection; in
other words, the Bargmann spacetime is a Brinkmann spacetime
\cite{MR1512246} or, equivalently, a pp-wave
\cite{MR0143624}. Starting with Minkowski spacetime, the null
reduction along a parallel null vector field gives Galilei spacetime.
In \cite{Julia:1994bs}, the null vector field need not be parallel,
but only Killing, and hence the Lorentzian manifolds go beyond the
class of pp-waves. Later, Gibbons and Patricot \cite{Gibbons:2003rv}
extended these ideas to obtain the non-relativistic limits of
(anti)~de~Sitter spaces, which they called Newton--Hooke spacetimes,
as null reductions of homogeneous pp-waves. These (A)dS--Galilei
spacetimes can be interpreted as non-relativistic spacetimes with a
non-vanishing cosmological constant. More recently Bekaert and Morand
\cite{Bekaert:2015xua} studied conditions under which the null
reduction by proper and free, but not necessarily isometric, action of
the additive reals on a Lorentzian manifold results in a Galilean
spacetime. We shall see examples of both the Julia--Nicolai null
reduction and the more general null reduction of Bekaert--Morand in
this paper.

Before we start, a word of nomenclature. From now on, by the word
`spacetime' we shall always mean a reductive homogeneous spacetime.
Any reductive homogeneous space carries a~canonical invariant
connection (defined by the vanishing of the Nomizu map, as explained
in the current context, for example, in
\cite{Figueroa-OFarrill:2019sex}), as well as a number of homogeneous
tensor fields arising by the holonomy principle from any invariant
tensors of the linear isotropy representation at any chosen ``origin''
of the spacetime, as will be recalled below. By reduction we shall
mean first and foremost reduction of homogeneous spaces, but in some
cases this reduction also results directly in a reduction of the
additional structure: connection and/or homogeneous tensor fields.

The symmetric Galilean spacetimes -- namely, Galilei and (anti)
de~Sitter--Galilei -- are homogeneous spaces of the Galilei and
Newton--Hooke groups. They are known to arise as null reductions of
certain homogeneous pp-wave spacetimes and hence it may be expected
that the torsional Galilean spacetimes also admit such a description.
In this paper, we show that this expectation is correct and exhibit
explicit homogeneous pp-wave spacetimes whose null reductions agree
with all the spatially isotropic homogeneous Galilean spacetimes
in~\cite{Figueroa-OFarrill:2018ilb}, including the torsional cases. We
also show that these pp-waves are solutions of the Einstein field
equations sourced by pure radiation fields. For convenience, we
collect Galilean spacetimes in Table~\ref{tab:Galilean}, whose
notation we now briefly explain.

\begin{table}[h]
 {\centering

 \caption{Spatially isotropic homogeneous Galilean spacetimes ($n>2$).}\vspace{1mm}

 \label{tab:Galilean}
 \rowcolors{2}{blue!10}{white}
 \begin{tabular}{>{$}l<{$}|>{$}l<{$}|l}\toprule
 \multicolumn{1}{c|}{Spacetime} & \multicolumn{1}{c|}{Additional nonzero Lie brackets} & \multicolumn{1}{c}{Name}\\\midrule
 \zG & & Galilei\\
 \zdSG & [H,\P] = -\B & de~Sitter--Galilei (also $\ztdSG_{\gamma=-1}$)\\
 \zdSG_\gamma & [H,\P] = \gamma\B + (1+\gamma)\P & torsional de~Sitter--Galilei ($\gamma\in (-1,1)$) \\
 \zdSG_1 & [H,\P] = \B + 2 \P & torsional de~Sitter--Galilei ($\gamma=1$) \\
 \zAdSG & [H,\P] = \B & anti~de~Sitter--Galilei (also $\ztAdSG_{\chi=0}$)\\
 \ztAdSG_\chi & [H,\P] = \big(1+\chi^2\big) \B + 2\chi \P & torsional anti~de~Sitter--Galilei ($\chi>0$) \\
 \bottomrule
 \end{tabular}

 }

 \medskip

 \noindent
In this table we provide the nonzero Lie brackets in
 addition to $[\J,\J] = \J$, $[\J, \B] = \B$, $[\J,\P] = \P$ and
 $[\B,H]=\P$, in a basis where the stabiliser subalgebra $\h$ is
 spanned by $\langle \J, \B \rangle$.
\end{table}

A kinematical Lie algebra (in spatial dimension $n$) is a real Lie
algebra with $\so(n)$ rotations accompanied by two vectors (boosts and
spatial translations) and one scalar (temporal translation). More
concretely, $\g$ is spanned by $J_{ij} = -J_{ji}$, $B_i$, $P_i$ and
$H$, where $i,j = 1,\dots, n$ and the Lie brackets include the
following
\begin{gather}
 [J_{ij}, J_{k\ell}] = \delta_{jk} J_{i\ell} - \delta_{ik} J_{j\ell} - \delta_{j\ell} J_{ik} + \delta_{i\ell} J_{jk},\nonumber\\
 [J_{ij}, B_k] = \delta_{jk} B_i - \delta_{ik} B_j,\nonumber\\
 [J_{ij}, P_k] = \delta_{jk} P_i - \delta_{ik} P_j,\nonumber\\
 [J_{ij}, H] = 0, \label{eq:kla-generic}
 \end{gather}
and any other brackets consistent with the Jacobi identities. It is
convenient to use an abbreviated notation in which we avoid the
indices and write the above brackets as
\begin{equation*} 
 [\J, \J] = \J,\qquad
 [\J, \B] = \B,\qquad
 [\J, \P] = \P,\qquad
 [\J, H] = 0.
\end{equation*}
In this work we focus on homogeneous kinematical spacetimes; that is,
homogeneous spacetimes of a kinematical Lie group $\Ggr$. Every such
spacetime is described infinitesimally by a Klein
pair\footnote{Originally, and unwisely, called a Lie pair in
 \cite{Figueroa-OFarrill:2018ilb}.} $(\g,\h)$, where $\g$, the Lie
algebra of $\Ggr$, is the transitive Lie algebra, and $\h$, the
isotropy subalgebra, is the Lie algebra of the subgroup $\Hgr$ of
$\Ggr$ which fixes a chosen origin for the homogeneous spacetime. It
is this choice of homogeneous spacetime that provides a physical
interpretation to the abstract Lie group $\Ggr$ and gives meaning to
what we call a boost (whose generator belongs to $\h$ along with the
rotations) and what we call a translation (which is in the
complement). For example, the Poincar\'e group gives rise to a plethora
of interesting inequivalent homogeneous spacetimes with different
physical interpretations, see, e.g.,~\cite{Figueroa-OFarrill:2021sxz}.

In Table~\ref{tab:Galilean}, we have chosen a basis for $\g$ in such a
way that the Lie subalgebra $\h$ is spanned by $J_{ij}$ and $B_i$, the
rotations and (Galilean) boosts. Every row in the table lists the
additional (i.e., not included in equation~\eqref{eq:kla-generic} or
$[\B,H]=\P$, which is common to Galilean transitive Lie algebras)
nonzero Lie brackets in such a basis in the above abbreviated
notation. As shown in~\cite{Figueroa-OFarrill:2018ilb}, the Klein
pairs in the table are in bijective correspondence with
(isomorphism classes of) simply-connected spatially isotropic
homogeneous Galilean spacetimes. These spacetimes form two continua
$\zdSG_\gamma$, for $-1\leq \gamma < 1$, and $\zAdSG_\chi$, for $\chi
\geq 0$. They both have a common limit $\lim_{\gamma \to 1}
\zdSG_\gamma = \lim_{\chi \to \infty} \zAdSG_\chi$, which is no
contraction and is represented by the red dot in
Figure~\ref{fig:Galilean}. All torsional Galilean spacetimes share a
common geometric limit (i.e., a contraction at the level of the
kinematical Lie algebras) to the Galilei spacetime $\zG$, as
illustrated by the blue arrows that leave the green arrows in
Figure~\ref{fig:Galilean}. That figure also contains the maximally
symmetric Lorentzian spacetimes and their Galilean (non-relativistic)
and zero-curvature limits.

The spacetime denoted $\zdSG_1$ is more properly thought of as a limit
$\lim_{\gamma \to 1} \zdSG_\gamma$ and it coincides with the limit
$\lim_{\chi\to\infty} \zAdSG_\chi$. All Galilean spacetimes are
reductive and $\zG$, $\zdSG$ and $\zAdSG$ are symmetric. They are the
ones which can be obtained as non-relativistic limits of Minkowski,
de~Sitter and anti~de~Sitter spacetimes, respectively.

This note is organised as follows. In
Section~\ref{sec:Lorentzian-klein-pairs}, we introduce some Klein
pairs corresponding to homogeneous pp-wave spacetimes; that is,
homogeneous Lorentzian manifolds admitting a~parallel (relative to the
Levi-Civita connection) null vector field. In
Section~\ref{sec:metrics}, we describe the invariant Lorentzian
metrics relative to (modified) exponential coordinates. In
Section~\ref{sec:riemann-ricci-tensor}, we calculate the curvature of
the pp-wave metrics and show that they are solutions of the Einstein
field equations sourced by pure radiation fields. In
Section~\ref{sec:killing-vectors}, we exhibit explicit expressions for
the Killing vector fields. In Section~\ref{sec:null-reductions}, we
discuss the null reductions along the parallel null vector and show
that they give the homogeneous Galilean spacetimes described in the
Introduction. The Galilean structure induced by the reduction only
partially agrees with the invariant Galilean structure: the spatial
cometric is the invariant one, but the clock one-form is only
homothetic to the invariant one; that is conformal to it but with a
constant conformal factor. We then introduce a different null
reduction resulting in the same homogeneous Galilean spacetime, but
inducing the invariant Galilean structure on the nose. Two of the
pp-wave metrics in Section~\ref{sec:metrics} are flat and hence
locally isometric, but their reductions result in non-isomorphic
homogeneous Galilean spacetimes. This is discussed in
Section~\ref{sec:GvdSG0}. Finally, in Section~\ref{sec:conclusions}, we
offer some conclusions. In Appendix~\ref{sec:global}, we provide a
complementary set of coordinates and study global properties of the
homogeneous pp-waves.

\section{Lorentzian Klein pairs of the pp-waves}
\label{sec:Lorentzian-klein-pairs}

In this section, we introduce a number of Klein pairs corresponding to
homogeneous Lorentzian manifolds admitting a parallel null vector
field; that is, homogeneous pp-wave spacetimes. We will later show
that their associated null reductions give all the homogeneous
Galilean spacetimes in Table~\ref{tab:Galilean}.

Before describing them, it is perhaps useful to say something about
where they come from. They arose initially in a forthcoming follow-up
paper to \cite{Figueroa-OFarrill:2022kcd} in which we discuss
geometries associated to Lie algebras obtained from Lifshitz Lie
algebras by the addition of boosts. The homogeneous pp-wave spacetimes
in question are geometric realisations of effective Klein pairs
$(\g,\h)$ where $\g$ is a deformation of the centrally extended static
kinematical Lie algebra and $\h$ is spanned by what could be
interpreted as spatial rotations and boosts. The Lie algebras $\g$ had
already appeared in \cite[Table~2]{Figueroa-OFarrill:2017ycu} for
spatial dimension $n=3$ and in
\cite[Table~18]{Figueroa-OFarrill:2017tcy} for $n>3$. These Lie
algebras also arise naturally in the description of particle dynamics
\cite{Figueroa-OFarrill:2022tlf} on the homogeneous Galilean
spacetimes in Table~\ref{tab:Galilean}.

Let us now describe the Lorentzian Klein pairs $(\g,\h)$ in question.
The Lie algebra $\g$ is spanned by $J_{ij} = -J_{ji}$, $B_i$, $P_i$,
$H$ and $Z$; although despite the notation $Z$ is not necessarily
central. The Lie brackets are written in the abbreviated form and we
list only the nonzero Lie brackets in~$\g$ in addition to $[\J,\J] =
\J$, $[\J, \B] = \B$, $[\J,\P] = \P$, $[\B,\P]=Z$ and $[\B,H]=\P$, in
a basis where the stabiliser subalgebra $\h$ is spanned by $J_{ij}$
and $B_i$. They are listed in Table~\ref{tab:Lorentzian-klein-pairs},
where the spacetimes have been labelled as $\mathsf{PX}$ with
$\mathsf{X}$ a Galilean spacetime, foreshadowing their interpretation
as a principal (right) $\RR$-bundle over $\mathsf{X}$.

\begin{table}[h]

{\centering
 \caption{Homogeneous Lorentzian spacetimes.} \label{tab:Lorentzian-klein-pairs}\vspace{1mm}

 \rowcolors{2}{blue!10}{white}
 \begin{tabular}{>{$}l<{$}|*{2}{>{$}l<{$}}|l}\toprule
 \multicolumn{1}{c|}{Spacetime} & \multicolumn{2}{c|}{Additional nonzero Lie brackets} & \multicolumn{1}{c}{Comments}\\\midrule
 \pG & & & \\
 \pdSG & [H,\P] = -\B & & \\
 \pdSG_\gamma & [H,\P] = \gamma\B + (1+\gamma)\P & [H,Z] = (1+\gamma) Z & $\gamma\in (-1,1)$ \\
 \pdSG_1 & [H,\P] = \B + 2 \P & [H,Z] = 2 Z & \\
 \pAdSG & [H,\P] = \B & & \\
 \ptAdSG_\chi & [H,\P] = \big(1+\chi^2\big) \B + 2\chi \P & [H,Z] = 2 \chi Z & $\chi>0$ \\
 \bottomrule
 \end{tabular}

 }

 \medskip

 \noindent
In this table we provide the nonzero Lie brackets in
 addition to $[\J,\J] = \J$, $[\J, \B] = \B$, $[\J,\P] = \P$,
 $[\B,\P]=Z$ and $[\B,H]=\P$, in a basis where the stabiliser
 subalgebra $\h$ is spanned by $\langle \J, \B \rangle$.
\end{table}

For each pair $(\g,\h)$ above we have a reductive decomposition
$\g = \m \oplus \h$ with $\m = \langle \P,H,Z\rangle$. Let
$\pi^a$, $\eta$, $\zeta$ be the canonically dual basis for $\m^*$, i.e.,
the nonzero relations are
$\langle \pi^{a}, P_{b}\rangle = \delta^{a}_{b}$,
$\langle \eta, H \rangle=1$ and $\langle Z, \zeta \rangle=1$. Then for
all Klein pairs in Table~\ref{tab:Lorentzian-klein-pairs}, there is an
$\Hgr$-invariant Lorentzian inner product on $\m$ given by
$\pi^2 - 2\eta\zeta$ and an $\Hgr$-invariant vector given by $Z$. The
holonomy principle associates to the inner product a $\Ggr$-invariant
Lorentzian metric $g$ on the homogeneous space with Klein pair $(\g,\h)$
and to the invariant vector a nowhere-vanishing null vector field
$\zeta$. Both the metric and the null vector field are parallel with
respect to the canonical invariant connection on the homogeneous
space, whose holonomy representation is the linear isotropy
representation of $\Hgr$. It is only in the case of a symmetric space
($[\m,\m] \subset \h$) that the canonical invariant connection agrees
with the Levi-Civita connection of $g$. In that
case, the null vector field $\zeta$ is parallel with respect to the
Levi-Civita connection and indeed agrees with the parallel null vector
of the pp-wave. In this case, and this case alone, is $\zeta$ also,
in particular, Killing.

The above reductive split lets us introduce the coset
parametrisation\footnote{We refer to $(u,v,\x)$ as modified
 exponential coordinates, to distinguish them from the truly
 exponential coordinates which would be defined by $\sigma(u,v,\x) =
 \exp(v Z + u H + \x \cdot \P)$.}
$\sigma\colon \m \to \Ggr$
\begin{equation}
 \label{eq:co-par}
 \sigma(u,v,\x) = \exp(v Z) \exp(\x \cdot \P) \exp(u H).
\end{equation}
This parametrisation is of course not unique and we provide a
complementary set of coordinates in Appendix~\ref{sec:global}. The
pull-back $\sigma^*\vartheta = \sigma^{-1} {\rm d}\sigma$ of the
Maurer--Cartan one-form on $\Ggr$ decomposes into
$\sigma^*\vartheta = \theta + \omega $, where $\theta$ is a local
coframe, in terms of which the $\Ggr$-invariant Lorentzian metric is
given by $g = \big(\pi^2 - 2 \eta\zeta\big)(\theta,\theta)$. The one-form
$\omega$ is the canonical invariant connection and it only agrees with
the Levi-Civita connection of the invariant metric in the symmetric
case; that is, when $[\m,\m] \subset \h$. This is only the case for
$\pG$, $\pdSG$ and $\pAdSG$. Not coincidentally, these are precisely
the cases where $Z$ is central and hence $\g$ is a central extension
of a Galilean kinematical Lie algebra: Bargmann in the case of $\pG$
and (anti)~de~Sitter--Bargmann in the cases of $\pdSG$ and $\pAdSG$
which are sometimes also called Bargmann--Newton--Hooke.

\section{The pp-wave metrics and the invariant vector fields}
\label{sec:metrics}

We now construct the explicit pp-wave metrics $g$ and the invariant vector
fields $\zeta$ in the modified exponential coordinates $\big(u,v,x^i\big)$ in
equation~\eqref{eq:co-par}. We calculate the coframe $\theta$ and
then we evaluate the metric $g = \big(\pi^2 - 2
\eta\zeta\big)(\theta,\theta)$. The value at $p \in M$ of the invariant
vector field $\zeta$ is the inverse image of $Z$ under the isomorphism
$T_pM \to \m$ defined by the coframe.

For $\pG$ we find that
\begin{equation*}
 \sigma^*\vartheta = {\rm d}u H + {\rm d}v Z + {\rm d}\x \cdot \P = \theta,
\end{equation*}
so that $\omega = 0$. We recognise the metric
\begin{equation}
 \label{eq:metric-G}
 g = {\rm d}\x\cdot {\rm d}\x - 2 {\rm d}u {\rm d}v
\end{equation}
as Minkowski spacetime in light-cone coordinates. This metric is of
course invariant under the full Poincar\'e group, but here we are
reducing the structure to the Bargmann subgroup singled out by the
distinguished null vector $\zeta = \d_v$ corresponding to $Z$.

For $\pdSG_\gamma$, we have
\begin{align*}
 \sigma^*\vartheta &= {\rm d}u H + \exp(-u \ad_H) ({\rm d}v Z + {\rm d}\x \cdot \P)\\
 &= {\rm d}u H + {\rm e}^{-u(1+\gamma)} {\rm d}v Z + (\exp(-u \ad_H) \P)\cdot {\rm d}\x.
 \end{align*}
From
\begin{equation*}
 -u \ad_H \begin{pmatrix} \B & \P \end{pmatrix}= \begin{pmatrix} \B &
 \P \end{pmatrix}
 \begin{pmatrix}
 0 & -\gamma u \\ u & -(1+\gamma) u
 \end{pmatrix}
\end{equation*}
we find that for $\gamma \in [-1,1)$,
\begin{equation*}
 \exp(-u \ad_H) \begin{pmatrix} \B & \P \end{pmatrix}= \begin{pmatrix} \B &
 \P \end{pmatrix}
 \frac1{1-\gamma}
 \begin{pmatrix}
 {\rm e}^{-u\gamma} - \gamma {\rm e}^{-u} & \gamma\big( {\rm e}^{-u} - {\rm e}^{-u\gamma} \big)\\
 {\rm e}^{-u\gamma} - {\rm e}^{-u} & {\rm e}^{-u}- \gamma {\rm e}^{-u\gamma}
 \end{pmatrix}
\end{equation*}
and for $\gamma = 1$,
\begin{equation*}
 \exp(-u \ad_H) \begin{pmatrix} \B & \P \end{pmatrix}= \begin{pmatrix} \B &
 \P \end{pmatrix}
 \begin{pmatrix}
 {\rm e}^{-u} (1+u) & - u {\rm e}^{-u} \\
 u {\rm e}^{-u} & {\rm e}^{-u}(1-u)\end{pmatrix},
\end{equation*}
from where we read off
\begin{equation*}
 (\exp(-u \ad_H) \P) =
 \begin{cases}
 \displaystyle \frac\gamma{1-\gamma} \left( {\rm e}^{-u} - {\rm e}^{-u\gamma} \right)\B + \frac1{1-\gamma}\left( {\rm e}^{-u} - \gamma {\rm e}^{-u\gamma} \right) \P, & \gamma \in [-1,1),\\
\displaystyle -u {\rm e}^{-u} \B + (1-u) {\rm e}^{-u}\P, & \gamma = 1.
 \end{cases}
\end{equation*}
The local coframe is then given by
\begin{equation} \label{eq:coframe-dSG}
 \theta = {\rm d}u H +
 \begin{cases}
\displaystyle {\rm e}^{-u(1+\gamma)} {\rm d}v Z + \frac1{1-\gamma}\big( {\rm e}^{-u} - \gamma {\rm e}^{-u\gamma} \big) \P\cdot {\rm d}\x, & \gamma \in [-1,1),\\
 {\rm e}^{-2u} {\rm d}v Z + (1-u) {\rm e}^{-u}\P \cdot {\rm d}\x,& \gamma = 1
 \end{cases}
\end{equation}
and the connection one form is given by
\begin{equation*}
 \omega =
 \begin{cases}
\displaystyle \frac\gamma{1-\gamma} \big( {\rm e}^{-u} - {\rm e}^{-u\gamma} \big)\B \cdot
 {\rm d}\x, & \gamma \in [-1,1),\\
 -u {\rm e}^{-u} \B \cdot {\rm d}\x & \gamma = 1 .
 \end{cases}
\end{equation*}
For $\gamma \leq 0$, the coframe is invertible for any $u \in \RR$
whereas for $\gamma \in (0,1)$ we have the restriction
$u < \frac{\ln \gamma}{\gamma-1}$. From
equation~\eqref{eq:coframe-dSG} we can determine the invariant vector
field $\zeta$ as well as the resulting metric $g$. With
$\gamma \in (-1,1)$ they are given by (we have singled out the two
extreme points):
\begin{equation*}
 \begin{split}
 (\pdSG) \quad &g= -2 {\rm d}u {\rm d}v + (\cosh u)^2 {\rm d}\x \cdot {\rm d}\x, \qquad
 \zeta = \d_v,\\
 (\pdSG_\gamma) \quad &g = -2 {\rm e}^{-u(1+\gamma)} {\rm d}u {\rm d}v +
 \frac1{(1-\gamma)^2} \big( {\rm e}^{-u} - \gamma {\rm e}^{-u\gamma} \big)^2
 {\rm d}\x \cdot {\rm d}\x, \qquad \zeta = {\rm e}^{(1+\gamma)u} \d_v,\\
 (\pdSG_1) \quad &g= -2 {\rm e}^{-2u} {\rm d}u {\rm d}v + {\rm e}^{-2u}(1-u)^2 {\rm d}\x \cdot
 {\rm d}\x, \qquad \zeta = {\rm e}^{2u}\d_v.
 \end{split}
\end{equation*}

Lastly, for $\pAdSG_\chi$ we have
\begin{align*}
 \sigma^*\vartheta &= {\rm d}u H + \exp(-u \ad_H) ({\rm d}v Z + {\rm d}\x \cdot \P)\\
 &= {\rm d}u H + {\rm e}^{-2u\chi} {\rm d}v Z + (\exp(-u \ad_H) \P)\cdot {\rm d}\x.
 \end{align*}
From
\begin{equation*}
 -u \ad_H \begin{pmatrix} \B & \P \end{pmatrix}= \begin{pmatrix} \B &
 \P \end{pmatrix}
 \begin{pmatrix}
 0 & -u \big(1+\chi^2\big) \\ u & -2 u\chi
 \end{pmatrix}
\end{equation*}
we find that
\begin{equation*}
 \exp(-u \ad_H) \begin{pmatrix} \B & \P \end{pmatrix}= \begin{pmatrix} \B &
 \P \end{pmatrix}
 {\rm e}^{-u\chi}
 \begin{pmatrix}
 \cos u + \chi \sin u & - \big(1+\chi^2\big) \sin u \\
 \sin u & \cos u - \chi \sin u,
 \end{pmatrix}
\end{equation*}
from where we read off
\begin{equation*}
 (\exp(-u \ad_H) \P) = {\rm e}^{-u\chi} (\cos u - \chi \sin u) \P -
 {\rm e}^{-u\chi} \big(1+ \chi^2\big) \sin u \B.
\end{equation*}
The local coframe is given by
\begin{equation*}
 \theta = {\rm d}u H + {\rm e}^{-2u\chi} {\rm d}v Z + {\rm e}^{-u\chi} (\cos u - \chi \sin
 u) \P \cdot {\rm d}\x
\end{equation*}
and the connection one-form is given by
\begin{equation*}
 \omega = -
 {\rm e}^{-u\chi} \big(1+ \chi^2\big) \sin u \B \cdot {\rm d}\x.
\end{equation*}
The invariant metrics and vector fields, where now $\chi > 0$ and we have
singled out the case $\chi = 0$, are given by
\begin{equation*}
 \begin{split}
 (\pAdSG) \quad &g= -2 {\rm d}u {\rm d}v + (\cos u)^2 {\rm d}\x \cdot {\rm d}\x, \qquad
 \zeta = \d_v\\
 (\pAdSG_\chi) \quad &g = -2 {\rm e}^{-2u\chi} {\rm d}u {\rm d}v + {\rm e}^{-2u\chi}
 ( \cos u - \chi \sin u )^2 {\rm d}\x \cdot {\rm d}\x, \qquad \zeta =
 {\rm e}^{2\chi u} \d_v.
 \end{split}
\end{equation*}
For $\chi >0$ the coframe is invertible in the range $-\frac\pi2 < u <
\arctan\big(\frac1\chi\big)$ and for $\chi =0$ in $-\pi/2 < u < \pi/2$.

In Appendix~\ref{sec:global}, using a different set of (global)
coordinates, we show that the simply-connected Lorentzian spacetimes
are actually diffeomorphic to $\RR^{n+2}$.

We summarise the results of this section in Table~\ref{tab:metrics}.

\begin{table}[h!]
 \centering
 \caption{Invariant metrics and vector fields.} \label{tab:metrics}\vspace{1mm}
 \rowcolors{2}{blue!10}{white}
 \setlength\extrarowheight{5pt}
 \begin{tabular}{*{3}{>{$}l<{$}|}l}\toprule
 \multicolumn{1}{c|}{Spacetime} & \multicolumn{1}{c|}{Metric $g$} & \multicolumn{1}{c|}{Vector field $\zeta$} & \multicolumn{1}{c}{Comments}\\\midrule
 \pG & -2 {\rm d}u {\rm d}v + d\x \cdot {\rm d}\x & \d_v & \\
 \pdSG & -2 {\rm d}u {\rm d}v + (\cosh u)^2 {\rm d}\x \cdot {\rm d}\x & \d_v & $\pdSG_{\gamma=-1}$ \\
 \pdSG_\gamma & -2 {\rm e}^{-(1+\gamma)u} {\rm d}u {\rm d}v + \big( \frac{{\rm e}^{-u} - \gamma {\rm e}^{- \gamma u}}{1-\gamma} \big)^{2} {\rm d}\x \cdot {\rm d}\x & {\rm e}^{(1+\gamma)u}\d_v & $\gamma\in (-1,1)$ \\
 \pdSG_1 & -2 {\rm e}^{-2u} {\rm d}u {\rm d}v + {\rm e}^{-2u} (1-u)^2 {\rm d}\x \cdot {\rm d}\x & {\rm e}^{2u}\d_v & \\
 \pAdSG & -2 {\rm d}u {\rm d}v + (\cos u)^2 {\rm d}\x \cdot {\rm d}\x & \d_v & $\ptAdSG_{\chi=0}$ \\
 \ptAdSG_\chi & -2{\rm e}^{-2\chi u} {\rm d}u {\rm d}v + {\rm e}^{-2\chi u} (\cos u - \chi \sin u)^2 {\rm d}\x \cdot {\rm d}\x & {\rm e}^{2\chi u}\d_v & $\chi>0$ \\
 \bottomrule
 \end{tabular}
\end{table}

\section{Curvature tensors}\label{sec:riemann-ricci-tensor}

Every metric in Table~\ref{tab:metrics} is of the following form
\begin{equation*}
 g = -2 a(u) {\rm d}u {\rm d}v + b(u) {\rm d}\x \cdot {\rm d}\x ,
\end{equation*}
for some functions $a(u)$ and $b(u)$, which can easily be read off.
Our first observation is that such a metric is conformally flat.
Indeed, away from the set of points where $b(u)=0$, we may factor out
$b(u)$ and write
\begin{equation*}
 g = b(u) \bigl( -2 \tfrac{a(u)}{b(u)} {\rm d}u {\rm d}v + {\rm d}\x \cdot {\rm d}\x\bigr)
\end{equation*}
and then simply define a new coordinate $t$ by ${\rm d}t=-a(u)/b(u) {\rm d}u$
so that the metric becomes
\begin{align*}
 g=b(u) (2 {\rm d}t {\rm d}v + {\rm d}\x \cdot {\rm d}\x),
\end{align*}
which is manifestly conformally flat. The Weyl tensor
vanishes outside the set of points where $b(u)=0$, which for the
metrics in Table~\ref{tab:metrics} is either the whole space (when~$b$
never vanishes, as in $\pG$ and $\pdSG$) or a dense open subset (in
the remaining metrics). In either case, if the Weyl tensor vanishes
in an open dense subset, it vanishes everywhere.

The nonzero components of the Riemann and Ricci tensor of the
Levi-Civita connection of $g$ are given by
\begin{gather*}
 R_{u x^i u x^i} = \tfrac{1}{4} \big( \tfrac{2 a' b'}{a} + \tfrac{b^{\prime 2}}{b} - 2 b'' \big),\\
 R_{uu} = \tfrac{n}{b} R_{u x^{i} u x^{i}}
\end{gather*}
and any other components related to these by the symmetries of the
Riemann tensor, where $i=1,\dots,n$ and where we do not sum over the
$i$ indices.

For the case at hand this reduces to
\begin{align*}
 (\pdSG) \quad &R_{u x^{i} u x^{i}}=-(\cosh u)^2,\qquad R_{uu}= -n,\\
 (\pdSG_\gamma) \quad &R_{u x^{i} u x^{i}}=\frac{\gamma}{(1-\gamma)^2} \big( {\rm e}^{-u} - \gamma {\rm e}^{-u\gamma} \big)^2 ,\qquad R_{uu} = n \gamma, \\
 (\pdSG_1) \quad & R_{u x^{i} u x^{i}}= {\rm e}^{-2u}(1-u)^2,\qquad R_{uu}= n
\end{align*}
and
\begin{align*}
 (\pAdSG) \quad &R_{u x^{i} u x^{i}}= (\cos u)^2,\qquad R_{uu}= n,\\
 (\pAdSG_\chi) \quad& R_{u x^{i} u x^{i}}=\big(1+\chi^{2}\big) {\rm e}^{-2u\chi} ( \cos u - \chi \sin u )^2,\qquad R_{uu}= n \big(1+\chi^{2}\big) .
\end{align*}
Notice that for $\gamma=0$, the Riemann tensor of the metric in
$\pdSG_{\gamma}$ vanishes and thus $\pdSG_{\gamma = 0}$ and $\pG$ are
locally isometric. We shall contrast these two cases in more detail
in Section~\ref{sec:GvdSG0}.

By inspection we also see that the Ricci scalar vanishes in all cases. This
means that the only non-vanishing components of the Einstein tensor are
the null Ricci tensor components $R_{uu}$. This suggests that these
metrics can be understood as solutions of the Einstein field equations
which are sourced by pure radiation fields (null dust) for which the
energy momentum tensor satisfies (see, e.g.,
\cite[Section~5.2]{Stephani:2003tm})
\begin{gather*}
 T_{\mu\nu}=\phi^2 k_{\mu}k_{\nu}, \qquad k_{\mu} k^{\mu} = 0 .
\end{gather*}
For our metrics this means that $k=du$ and $\phi^2=n\gamma$ or
$\phi^2=n\big(1+\chi^2\big)$, which would seem to require $\gamma \geq 0$.
The physical interpretation of the metrics $\pdSG_{\gamma < 0}$ is not
so clear.

\section{The Killing vectors}\label{sec:killing-vectors}

We will now exhibit explicit expressions for the Killing vector fields
of the metrics in Table~\ref{tab:metrics}. For each generator $X \in
\g$, we will exhibit vector fields $\xi_X$ such that $[\xi_X,\xi_Y] =
- \xi_{[X,Y]}$ for all $X,Y \in \g$.

In all cases $\xi_{J_{ij}} = -x_i \d_j + x_j \d_i$, $\xi_{P_i} = \d_i$
and $\xi_Z = \d_v$. It then remains to give expressions for
$\xi_{B_i}$ and $\xi_H$. We find that
\begin{gather}
 \xi_{B_i} = x_i \d_v + f(u) \d_i,\nonumber\\
 \xi_H = \d_u + h(u) x^i \d_i + \big(\lambda v + \tfrac12 \mu x^2\big) \d_v,\label{eq:H-B-KVFs}
 \end{gather}
for functions $f$, $h$ of $u$ and constants $\lambda$, $\mu$. The value of
$\lambda$ is determined from the $[\xi_H,\xi_Z]$ bracket and that of
$\mu$ by $[\xi_H, \xi_{P_i}]$, which also gives an algebraic relation
allowing us to solve for $h$ in terms of $f$. Finally, the bracket
$[\xi_{B_i},\xi_H]$ gives a first-order ODE for $f$, which we can
solve in each case. The results of these calculations are summarised
in Table~\ref{tab:killing-data}. As a check on the calculations, one
can show that the invariant vector fields in Table~\ref{tab:metrics}
commute with all the Killing vectors.

\begin{table}[h]
 \centering
 \caption{Data for $\xi_{B_i}$ and $\xi_H$ in equation~\eqref{eq:H-B-KVFs}.} \label{tab:killing-data}\vspace{1mm}

 \setlength{\extrarowheight}{2pt}
 \rowcolors{2}{blue!10}{white}
 \begin{tabular}{>{$}l<{$}*{4}{|>{$}l<{$}}}
 \multicolumn{1}{c|}{Spacetime} & \multicolumn{1}{c|}{$\lambda$} & \multicolumn{1}{c|}{$\mu$} & \multicolumn{1}{c|}{$f(u)$} & \multicolumn{1}{c}{$h(u)$}\\\toprule
 \pG & 0 & 0 & u & 0\\
 \pdSG & 0 & -1 & \tanh u& - \tanh u\\
 \pdSG_{\gamma\in(-1,1)} & 1+\gamma & \gamma & - \frac{{\rm e}^{-u} - {\rm e}^{-\gamma u}}{{\rm e}^{-u} - \gamma {\rm e}^{-\gamma u}} & \frac{{\rm e}^{-u} - \gamma^2 {\rm e}^{-\gamma u}}{{\rm e}^{-u} - \gamma {\rm e}^{-\gamma u}}\\
 \pdSG_1 & 2 & 1 & \frac{u}{1-u} & \frac{2-u}{1-u}\\
 \pAdSG & 0 & 1 & \tan u & \tan u\\
 \pAdSG_{\chi>0} & 2\chi & 1 + \chi^2 & \frac{\sin u}{\cos u - \chi \sin u} & 2\chi + \big(1+\chi^2\big)\frac{\sin u}{\cos u - \chi \sin u} \\\bottomrule
 \end{tabular}
\end{table}

\section{The null reductions}\label{sec:null-reductions}

We now show that the homogeneous pp-waves null reduce to the
torsional Galilean spaces as homogeneous spaces (spacetimes). We do
this first via Killing reduction, which is always guaranteed to result
in a Galilean structure in the quotient. Doing so, however, we find
that the reduced invariant structure matches the clock one-form of the
torsional spaces only up to scale. This is then remedied by
performing a reduction by the invariant vector field $\zeta$. This
results in the same homogeneous quotient (since $\zeta$ is invariant)
but now with the invariant Galilean structure.

\subsection{Killing reduction of spacetime}
\label{sec:reduction-spacetime}

Each of the Lorentzian metrics in Table~\ref{tab:metrics} possesses a
nowhere vanishing null Killing vector field $\xi = \d_v$ in the
modified exponential coordinates employed above. This Killing vector
field generates a one-parameter subgroup $\Gamma$ of the isometry
group of $(M,g)$: the one generated by $Z \in \g$. The space of orbits
$M/\Gamma$ can be given the structure of a smooth manifold in such a~way that the canonical projection $\pi\colon M \to M/\Gamma$ taking a
point to its orbit under $\Gamma$ is a smooth map. This allows us to
pull back functions and differential forms from $M/\Gamma$ to $M$ and
sets up an isomorphism of $C^\infty(M/\Gamma)$-modules between the
forms on $M/\Gamma$ and the basic forms on $M$, where we remind the
reader that basic forms are those forms $\alpha$ which are horizontal,
so that $\imath_\xi \alpha =0$, and invariant, so that $\eL_\xi \alpha
= 0$. Equivalently, $\alpha$ is basic if both $\imath_\xi \alpha =0$
and $\imath_\xi {\rm d}\alpha = 0$.

Let $\xi^\flat = g(\xi,-)$ be the one-form metrically dual to $\xi$.
Then $\imath_\xi \xi^\flat = g(\xi,\xi) = 0$ because $\xi$ is null and
$\eL_\xi \xi^\flat = (\eL_\xi \xi)^\flat = 0$, where we have used that
$\xi$ is a Killing vector field, so that it commutes with the musical
isomorphisms. This shows that $\xi^\flat = \pi^*\tau$, for some
one-form $\tau \in \Omega^1(M/\Gamma)$.

If $\alpha \in \Omega^1(M/\Gamma)$, we may construct a vector field
$(\pi^*\alpha)^\sharp$ on $M$ by pulling $\alpha$ back to $M$ and then
applying the musical isomorphism. Given
$\alpha,\beta \in \Omega^1(M/\Gamma)$ we get a function
\begin{equation*}
 g\big((\pi^*\alpha)^\sharp, (\pi^*\beta)^\sharp\big) =
 (\pi^*\alpha)\big((\pi^*\beta)^\sharp\big) = \pi^* \alpha\big(\pi_*
 (\pi^*\beta)^\sharp\big).
\end{equation*}
This defines $\psi \in \Gamma\bigl(\odot^2 T(M/\Gamma)\bigr)$ by
\begin{equation*}
 \psi(\alpha,\beta) = \alpha\big(\pi_* (\pi^*\beta)^\sharp\big).
\end{equation*}
Clearly, if $\beta = \tau$, then $(\pi^*\beta)^\sharp= \xi$ and since
$\pi_* \xi = 0$, we see that $\psi(\alpha,\tau)= 0$ for all
$\alpha \in \Omega^1(M)$. The pair $(\tau,\psi)$ defines a Galilean
structure on $M$. Galilean structures can be classified into several
types according to their intrinsic torsion ${\rm d}\tau$
\cite{Bekaert:2014bwa, Christensen:2013lma, Figueroa-OFarrill:2020gpr, MR334831}. In all examples in this section, the
null killing vector $\xi = \d_v$ is actually parallel, so that
${\rm d}\xi^\flat =0 $ and hence ${\rm d}\tau= 0$. So that the intrinsic torsion
of the Galilean spacetimes vanishes in all cases.

Below we will calculate local coordinate expressions for $\tau$ and
$\psi$, so let us unpack the previous discussion and re-express
everything in a local chart. It is often convenient to work in local
coordinates which are adapted to the reduction. This means choosing
local coordinates $x^\mu = (x^a,v)$ for $M$ such that $x^a$
are local coordinates for the base $M/\Gamma$. Since functions on the
base lift to $\Gamma$-invariant functions on $M$, it follows that $\xi
x^a = 0$, so that $\xi \propto \d_v$. It is convenient to choose the
coordinate $v$ to be adapted to $\xi$, so that $\xi = \d_v$.
Nevertheless we will write $\xi = \xi^\mu \d_\mu$, with the tacit
understanding that $\xi^a =0$ and $\xi^v = 1$. The metric has a
local expression $g = g_{\lambda\rho} {\rm d}x^\lambda {\rm d}x^\rho$ and $g_{vv}
= 0$ since $\xi$ is null. The dual one-form $\xi^\flat$ has a local
expression $\xi^\flat = \xi_\mu {\rm d}x^\mu$, where $\xi_\mu = g_{\mu\rho} \xi^\rho$.
Since $\xi$ is null, it follows that $\xi_v = 0$ and the only
nonzero components are $\xi_a =: \tau_a$, the clock
one-form. It is locally a one-form on $M/\Gamma$ since $\xi$ is
also a Killing vector. To obtain a~local coordinate expression for
$\psi$, let $\alpha$, $\beta$ be two one-forms on the base, whose local
expressions are $\alpha = \alpha_a {\rm d}x^a$ and similarly for
$\beta$. Then $\psi(\alpha,\beta) = \psi^{ab} \alpha_a\beta_b =
g_{\lambda\rho} g^{\lambda a} g^{\rho b} \alpha_a \beta_b$, so that
$\psi^{ab} = g^{ab}$. Below we will actually choose local coordinates
$x^a = \big(u, x^i\big)$ and we will see that the only nonzero components of
$\psi$ are $\psi^{ij}$ and that~$\tau$ is proportional to~${\rm d}u$.

Let $(\g,\h)$ be one of the pp-wave Klein pairs in
Table~\ref{tab:Lorentzian-klein-pairs}. It is clear that in all cases,
the generator $Z \in \g$ spans an ideal $\langle Z\rangle$ of $\g$.
Quotienting by the action of the one-parameter subgroup generated by
$Z$ gives rise to a homogeneous space with Klein pair
$\big(\overline\g,\overline\h\big)$, where $\overline\g = \g/\langle Z\rangle$
and $\overline\h = \h/\h\cap\langle Z\rangle \cong \h$. It is clear by
inspection of Table~\ref{tab:Lorentzian-klein-pairs} that for each
such Klein pair $(\g,\h)$, the quotient Klein pair
$\big(\overline\g,\overline\h\big)$ is the corresponding one in
Table~\ref{tab:Galilean}. In other words, the homogeneous pp-wave
spacetimes in Table~\ref{tab:Lorentzian-klein-pairs} can be realised
as the total space of principal (right) $\RR$-bundles over the
homogeneous Galilean spacetimes in Table~\ref{tab:Galilean}. In
summary, this exhibits the Galilean spatially isotropic homogeneous
spacetimes in Table~\ref{tab:Galilean} as null reductions of
homogeneous pp-wave spacetimes with metrics given by
Table~\ref{tab:metrics}.

\subsection{Reduction of invariant structure}\label{sec:reduct-invar-struct}

A natural question is whether we recover them not just as homogeneous
spaces, but whether the induced Galilean structure is the invariant
one. The clock one-form $\tau$ pulls back to the one-form metrically
dual to $\xi_Z$. It follows that if $X \in \g$ does not commute
with $Z$, the fundamental vector field $\overline\xi_X$ in the
quotient does not preserve $\tau$. Indeed, we have that
\begin{equation*}
 \pi^* \eL_{\overline\xi_X} \tau = \eL_{\xi_X} \pi^*\tau =
 \eL_{\xi_X} \xi_Z^\flat = [\xi_X,\xi_Z]^\flat = \xi_{[Z,X]}^\flat.
\end{equation*}
It is clear from the Lie brackets in
Table~\ref{tab:Lorentzian-klein-pairs}, that $H$ is the only generator
of $\g$ which may fail to commute with $Z$ and, when this happens, it
simply rescales $Z$ by a constant. Suppose that $[Z,H] = w Z$ for some
weight $w$. Then it follows from the above calculation that
$\eL_{\overline\xi_H} \tau = w \tau$, so that it acts homothetically
on the clock one-form.

It is a simple matter to read off the Galilean structure $(\tau,
\psi)$ relative to the coordinates $\big(t=-u, x^i\big)$ for each of the above
spacetimes. These are listed in Table~\ref{tab:null-reds}. That table
also lists the homogeneous Galilean spacetimes corresponding to the
null reductions, using the notation of~\cite{Figueroa-OFarrill:2018ilb}. They cannot be immediately compared
with the ones in~\cite{Figueroa-OFarrill:2019sex} since we are using a
different coordinate system, but they can be recognised by their Klein
pair as explained above.

\begin{table}[h!]
 \centering
 \caption{Galilean structures of null reductions.} \label{tab:null-reds}\vspace{1mm}

 \setlength{\extrarowheight}{2pt}
 \rowcolors{2}{blue!10}{white}
 \begin{tabular}{>{$}l<{$}*{2}{|>{$}l<{$}}}
 \multicolumn{1}{c|}{Spacetime} & \multicolumn{1}{c|}{$\tau$} & \multicolumn{1}{c|}{$\psi$}\\\toprule
 \zG & {\rm d}t & \delta^{ij}\d_i \otimes \d_j\\
 \zdSG & {\rm d}t& (\cosh t)^{-2} \delta^{ij}\d_i \otimes \d_j\\
 \zdSG_\gamma & {\rm e}^{(1+\gamma)t} {\rm d}t & \big(\frac{{\rm e}^t - \gamma {\rm e}^{\gamma t}}{1-\gamma} \big)^{-2} \delta^{ij}\d_i \otimes \d_j\\
 \zdSG_{1} & {\rm e}^{2t}{\rm d}t & {\rm e}^{-2t}(1+t)^{-2} \delta^{ij} \d_i \otimes \d_j\\
 \zAdSG & {\rm d}t & (\cos t)^{-2} \delta^{ij}\d_i \otimes \d_j\\
 \zAdSG_\chi & {\rm e}^{2\chi t} {\rm d} t & {\rm e}^{-2\chi t} (\cos t + \chi \sin t)^{-2} \delta^{ij} \d_i \otimes \d_j\\\bottomrule
 \end{tabular}
\end{table}

The fundamental vector fields $\overline\xi_X$ in the quotient are
easy to determine from the expressions of the Killing vector fields:
all we need to do is drop any component along $\xi_Z = \d_v$. Doing
so, we see that $\overline\xi_{J_{ij}}$ and $\overline\xi_{P_i}$ are
given formally by the same expression as the Killing vector fields,
whereas $\overline\xi_{B_i} = f(u) \d_i$ and $\overline\xi_H = \d_u +
h(u) x^i \d_i$, where~$f(u)$ and~$h(u)$ can be read off from
Table~\ref{tab:killing-data}. It is easy to check that these
fundamental vector fields define an anti-representation of the
Galilean kinematical Lie algebra $\overline\g$. It follows from an
explicit calculation that the Galilean structure $(\tau,\psi)$ in the
quotient is invariant under the first derived ideal
$[\overline\g,\overline\g]$, which is spanned by $J_{ij}$, $B_i$, $P_i$.
On the other hand, the generator $H$ leaves invariant $\psi$, but acts
homothetically on~$\tau$, as explained above.

Indeed, it is a simple calculation to check that for $\zG$, $\zdSG$
and $\zAdSG$, the Galilean structure induced from the null reduction
is the invariant one, whereas for $\zdSG_\gamma$, one finds
$\eL_{\overline\xi_H} \tau = (1+\gamma)\tau$ and for $\zAdSG_\chi$,
one finds $\eL_{\overline\xi_H} \tau = 2\chi \tau$. It follows that
the invariant clock one-form is conformal to the one induced by the
null-reduction. Indeed, for $\zdSG_\gamma$, it is
${\rm e}^{-(1+\gamma)t} \tau$ which is invariant, whereas for $\zAdSG_\chi$,
the invariant clock one-form is ${\rm e}^{-2\chi t}\tau$. In
\cite[Appendix~D]{Figueroa-OFarrill:2022tlf}, it is shown that there
exists a conformal rescaling of the Lorentzian metric such that the
vector field $\xi_Z$ remains Killing and the corresponding null
reduction does give the invariant clock-one form on the nose; although
one now pays the price that the induced spatial cometric is only
homothetic to the invariant one.

\subsection{Reduction by invariant vector field}\label{sec:an-improved-null}

We may solve the above problems by reducing via the action of the
reals generated by the invariant vector field $\zeta$. This vector
field is parallel with respect to the canonical invariant connection,
which is compatible with the invariant metric $g$, but has typically
nonzero torsion. It is therefore not Killing in the torsional cases,
but since it commutes with Killing vector fields, its flow commutes
with the $G$-action. This means that the quotient $M/\Gamma_\zeta$,
with $\Gamma_\zeta$ the one-dimensional subgroup of diffeomorphisms of
$M$ generated by $\zeta$, admits an action of $G$. However the normal
subgroup $\Gamma \subset G$ generated by $Z$ acts trivially on
$M/\Gamma_\zeta$. Indeed, the unparametrized integral curves of
$\zeta$ and $\xi_Z$ agree. To see this, notice from
Table~\ref{tab:metrics}, that $\zeta = {\rm e}^{\beta u} \d_v$ in all cases,
for some $\beta\in \RR$. The integral curves of $\zeta$ are given by
\begin{equation*}
 c(s) = \big(u_0, v_0 + {\rm e}^{\beta u_0} s, \x_0\big),
\end{equation*}
whereas the integral curves of $\xi_Z = \d_v$ are given by
\begin{equation*}
 c(s) = (u_0, v_0 + s, \x_0).
\end{equation*}

As a consequence, $\overline M := M/\Gamma_\zeta$ is a homogeneous
space of $\overline G = G/\Gamma$ and provides a geometric
realisation of the Klein pair $\big(\overline\g,\overline\h\big)$, where
$\overline\g = \g/\langle Z\rangle$ and
$\overline\h = \h/\h\cap\langle Z\rangle \cong \h$, as above.

One can check that the induced Galilean structure agrees with
the invariant one, but there is no need to do this explicitly, since
the fact that the action of $\Gamma_\zeta$ on $M$ commutes with
the $G$-action, any projectable $G$-invariant tensor field on $M$
(i.e., one descending to $\overline M$) is automatically $\overline
G$-invariant.

For example, it follows from the explicit expressions of $\zeta$ in
Table~\ref{tab:metrics}, that the metrically dual one-form
$\zeta^\beta$ is $-{\rm d}u$ in all cases, which agrees with
the invariant clock one-forms on the Galilean spacetimes in the
coordinates $\big(t=-u,x^i\big)$, as can be checked from the formulae in
Section~\ref{sec:metrics}, but ignoring the $Z$-component of the
soldering forms. The expressions for the induced spatial cometric
$\psi$ are precisely those in Table~\ref{tab:null-reds}.

\section[Is G = dSG\_0?]{Is $\boldsymbol{\zG = \zdSG_0}$?}
\label{sec:GvdSG0}

As we saw above, the invariant Lorentzian metric in $\pdSG_{\gamma = 0}$
is flat. This means that the spacetime $\pdSG_{\gamma=0}$ is (perhaps
only a portion of) Minkowski spacetime. The coordinates~$u$,~$v$,~$\x$ are
unconstrained, but they are not flat coordinates. Indeed, relative to
these coordinates, the metric takes the form
\begin{equation*}
 g = - 2 {\rm e}^{-u} {\rm d}u {\rm d}v + {\rm e}^{-2u} {\rm d}\x \cdot {\rm d}\x.
\end{equation*}
This suggests first of all defining a new coordinate $t = {\rm e}^{-u}$,
which takes only positive real values. Relative to $\big(t,v,x^i\big)$ the
metric becomes
\begin{equation*}
 g = 2 {\rm d}t {\rm d}v + t^2 {\rm d}\x \cdot {\rm d}\x.
\end{equation*}
We wish to find flat coordinates for this metric. Flat coordinates
are characterised by the fact that the coordinate vector fields are
parallel and the metrically dual one-forms are exact. Hence to find
them, we must first determine the parallel vectors. A vector field $\xi =
\xi^\mu \d_\mu$ is parallel if
\begin{equation*}
 \d_\mu \xi^\rho + \Gamma_{\mu\nu}{}^\rho \xi^\nu = 0.
\end{equation*}
We may find the Christoffel symbols by comparing the Euler--Lagrange
equations of the lagrangian
\begin{equation*}
 L = \dot t \dot v + \tfrac12 t^2 \dot\x \cdot \dot\x
\end{equation*}
with the geodesic equation
\begin{equation*}
 \ddot x^\rho + \Gamma_{\mu\nu}{}^\rho \dot x^\mu \dot x^\nu = 0.
\end{equation*}
Doing so, we find the following nonzero Christoffel symbols
\begin{equation*}
 \Gamma_{i j}{}^v = -t \delta_{ij} \qquad\text{and}\qquad
 \Gamma_{t i}{}^j = \Gamma_{i t}{}^j = \tfrac1t \delta^j_i.
\end{equation*}
The vector field $\xi = \xi^\mu \d_\mu$ is parallel if and only if the
components $\xi^\mu$ satisfy the following partial differential
relations:
\begin{alignat*}{4}
 & \d_\mu \xi^t = 0, \qquad && \d_t \xi^i + \tfrac1t \xi^i = 0, \qquad && \d_t \xi^v = 0, &\\
& \d_v \xi^\mu = 0, \qquad && \d_i \xi^j + \tfrac1t \delta_i^j \xi^t = 0,\qquad && \d_i \xi^v - t \delta_{ij} \xi^j = 0.&
 \end{alignat*}
Solving these relations, we find the following parallel vector fields
\begin{equation*}
 \d_v, \qquad x^i \d_v + \tfrac1t \d_a, \qquad \d_t - \tfrac1t x^i \d_i - \tfrac12 x^2 \d_v.
\end{equation*}
The metrically dual one-forms are, respectively,
\begin{equation*}
 {\rm d}t, \qquad {\rm d}\big(t x^i\big) \qquad\text{and}\qquad {\rm d}\big(v-\tfrac12 t x^2\big),
\end{equation*}
from where we can read off the flat coordinates
\begin{equation*}
 T = t, \qquad X^i = t x^i \qquad\text{and}\qquad V = v - \tfrac12 t x^2,
\end{equation*}
which we may invert to write
\begin{equation*}
 t = T, \qquad x^i = \tfrac{X^i}{T} \qquad\text{and}\qquad v = V + \tfrac12 \tfrac{X^2}T.
\end{equation*}
Expressing the metric relative to these coordinates, we find
\begin{equation*}
 g = 2 {\rm d}T {\rm d}V + {\rm d}\X \cdot {\rm d} \X,
\end{equation*}
which is indeed manifestly flat. Notice that since $t>0$, it is also
the case that $T>0$ so the original metric covers half of Minkowski
spacetime and the new coordinates allow us to extend the metric to the
whole of Minkowski spacetime.

The null vector $\xi_Z = \d_v$ becomes
\begin{equation*}
 \d_v = \frac{\d T}{\d v} \d_T + \frac{\d V}{\d v} \d_V + \frac{\d X^i}{\d v} \d_{X^i} = \d_V
\end{equation*}
in the new coordinates. The null reduction of this metric along
$\d_V$ is clearly the same as the null reduction of the metric in
equation~\eqref{eq:metric-G} along $\d_v$: simply change coordinates
to $t = -u$ and notice that the metrics agree under $\big(t,v,x^i\big) \mapsto
\big(T,V,X^i\big)$.

This does not mean, however, that the homogeneous Galilean spacetimes
$\zdSG_0$ and $\zG$ are the same. One way to see this is to notice
that under the improved reduction by the invariant vector field
$\zeta$, which results in the invariant Galilean structure, the
reductions are not the same. For $\zG$, it is still the case that
$\zeta = \d_v$, but for $\zdSG_0$, $\zeta = {\rm e}^u \d_v$ or, after the
change of coordinates, $\zeta = \frac1{T}\d_V$, which does not extend
smoothly to the full Minkowski spacetime in $(T,V,\X)$ coordinates.

\section{Discussion and conclusions}\label{sec:conclusions}

In this note we have exhibited the spatially isotropic homogeneous
Galilean spacetimes (of spatial dimension $n>2$) classified in
\cite{Figueroa-OFarrill:2018ilb} as null reductions of homogeneous
pp-wave spacetimes. We have performed two null reductions with
identical quotients as homogeneous spacetimes: one by a Killing vector
field and one by an invariant vector field. These two reductions
agree for the non-torsional (i.e., symmetric) homogeneous spacetimes,
but are different for the torsional ones. In the former reduction, we
find that although the null reduction gives the homogeneous Galilean
spacetimes as homogeneous spaces, the Galilean structure induced from
the null reduction is not the invariant one in the torsional cases
($\zdSG_\gamma$ for $-1 < \gamma \leq 1$ and $\zAdSG_\chi$ for $\chi >
0$), but only homothetic to it. This is a reflection of the fact that
the null Killing vector is not central in the torsional cases. In
contrast, the null reduction by the invariant vector field results in
the invariant Galilean structure by design.

This work was focused on Galilean spacetimes and their intricate
properties. As final note let us advertise their Carrollian curved
counterparts (A)dS--Carroll, which arise as a ultra-relativistic limit
of (A)dS. They are complementary equally interesting candidates for
the study of holography in a possibly more tractable setup (besides
the intriguing relation of AdS-Carroll to time-like
infinity~\cite{Figueroa-OFarrill:2021sxz}). They can be seen as
null hypersurfaces of (A)dS~\cite{Figueroa-OFarrill:2018ilb} and
AdS--Carroll shares the box-like (its spatial metric is hyperbolic)
and dS--Carroll the cosmological character (its spatial metric is the
sphere) of their Lorentzian parents~\cite{Figueroa-OFarrill:2019sex}.

\appendix

\section{Global properties of the Lorentzian spacetimes}
\label{sec:global}

In this appendix we show that every simply-connected homogeneous
Lorentzian spacetime $M$ whose Klein pair is listed in
Table~\ref{tab:Lorentzian-klein-pairs} is diffeomorphic to
$\RR^{n+2}$.

To do this we introduce another set of coordinates $(u,v,x^a)$ and
define the parametrisation
\begin{equation*}
 \sigma(u,v,\x) = \exp(v Z) \exp(u H) \exp(\x \cdot \P) ,
\end{equation*}
which is closer in spirit to the one used in~\cite[Appendix
A.2]{Figueroa-OFarrill:2019sex} for the Galilean spacetimes. With
$[H,\P] = \alpha \B +\beta \P$ and $[H,Z] = \beta Z$ and using
\begin{equation*}
 {\rm e}^{- \x \cdot \P} H {\rm e}^{\x \cdot \P} = H + \alpha \x \cdot \B +
 \beta \x \cdot \P + \tfrac12 \alpha x^2 Z,
\end{equation*}
we obtain
\begin{equation}
 \label{eq:coord2}
 \sigma^*\vartheta = {\rm d}u H + \big({\rm e}^{- \beta u} {\rm d}v + \tfrac12 \alpha
 x^2 {\rm d}u \big) Z + ({\rm d}\x + \beta {\rm d}u \x) \cdot \P+ \alpha {\rm d}u \x
 \cdot \B .
\end{equation}
The soldering form
\begin{equation*}
 \theta = {\rm d}u H + \big({\rm e}^{- \beta u} {\rm d}v + \tfrac12 \alpha
 x^2 {\rm d}u \big) Z + ({\rm d}\x + \beta {\rm d}u \x) \cdot \P
\end{equation*}
is everywhere invertible.

Let us define a map $j\colon \RR^{n+2} \to M$ by $j(u,v,\x) =
\sigma(u,v,\x)\cdot o$, where $o \in M$ is a choice of origin with
stabiliser the subgroup $\Hgr$ generated by spatial rotations and
boosts. The first thing to remark is that we may ignore the rotations
and think of $M$ simply as a homogeneous space of the solvable Lie
group $\Kgr$ generated by $\B$, $\P$, $H$, $Z$. The stabiliser of the origin
is then the subgroup $\Bgr$ spanned by~$\B$. We define an action of~$\Kgr$ on~$\RR^{n+2}$ by requiring~$j$ to be equivariant. Introducing
the shorthands
\begin{equation*}
 \varpi := \tfrac12 \sqrt{4\alpha - \beta^2 }, \qquad s :=
 \frac{\sin(\varpi u)}{2\varpi} \qquad\text{and}\qquad f_{\pm} = \cos(\varpi u) \pm \beta s,
\end{equation*}
a short calculation reveals that
\begin{gather*}
 {\rm e}^{a Z} \cdot (u,v,\x) = (u, v + a, \x),\\
 {\rm e}^{b H} \cdot (u,v,\x) = \big(u+b, {\rm e}^{b\beta} v, \x\big),\\
 {\rm e}^{\w \cdot \B} \cdot (u,v,\x) = \big(u, v + s f_+ w^2 + {\rm e}^{u\beta/2} f_+ \w \cdot \x, \x + 2 {\rm e}^{-u\beta/2} s \w \big),\\
 {\rm e}^{\y \cdot \P} \cdot (u,v,\x) = \big(u, v - \alpha s f_- y^2 - 2 \alpha u^{u\beta/2} s \y \cdot \x, \x + {\rm e}^{-u\beta/2} f_- \y\big),
 \end{gather*}
from where we may read the fundamental vector fields:
\begin{gather*}
 \xi_Z = \d_v,\\
 \xi_H = \d_u + \beta v \d_v,\\
 \xi_{B_a} = {\rm e}^{u\beta/2} f_+ x^a \d_v + 2 {\rm e}^{-u\beta/2} s \d_a,\\
 \xi_{P_a} = -2 \alpha {\rm e}^{u \beta/2} s x^a \d_v + {\rm e}^{-u\beta/2} f_- \d_a.
 \end{gather*}
One can check that these form an anti-representation of the Lie
algebra $\k$ of $\Kgr$.

Our first observation is that the action of $\Kgr$ on $\RR^{n+2}$ is
transitive. Indeed, taking as origin the point with coordinates
$(0,0,\mathbf{0})$, we see that we can reach any other point by the
action of $\sigma(u,v,\x) \in \Kgr$:
\begin{equation*}
 (u,v,\x) = \sigma(u,v,\x) \cdot (0,0,\mathbf{0}),
\end{equation*}
which follows essentially by definition. The smooth map
$j\colon \RR^{n+2} \to M$ is a local diffeomorphism, since the soldering
form is everywhere invertible, and it is $\Kgr$-equivariant. Together
with transitivity, this says that $j$ is a branched covering map, but
the branched locus has to be empty, otherwise it would not be
homogeneous. Therefore $j$ is a covering which, since
$\RR^{n+2}$ is simply connected, is universal. But $M$ was assumed to
be the simply-connected homogeneous spacetime realising the given
Klein pair, hence $j$ is an isomorphism of $\Kgr$-homogeneous
spacetimes and, in particular, $M$ is diffeomorphic to $\RR^{n+2}$.

These coordinates have other advantages. For example, the connection
one-form is uniformly given by $\omega = \alpha {\rm d}u \x \cdot \B$,
see the last term in~\eqref{eq:coord2}, and the invariant Lorentzian
metric $g$ and vector field $\zeta$ can be brought to a more uniform
form. To see this, let us write the invariant Lorentzian metric $g =
\big(\pi^2 - 2 \eta\zeta\big)(\theta,\theta)$ as
\begin{equation*}
 g=-2 \big({\rm e}^{- \beta u} {\rm d}v + \tfrac{\alpha}{2} x^{2} {\rm d}u \big){\rm d}u + ({\rm d}\x + \beta {\rm d}u \x)^{2} ,
\end{equation*}
and the invariant vector field $\zeta$ as
\begin{equation*}
 \zeta = {\rm e}^{\beta u} \d_v.
\end{equation*}

\subsection*{Acknowledgments}

We are grateful to Dieter Van den Bleeken, who spotted an erroneous
claim in an earlier version of this note. JMF would like to
acknowledge many useful discussions with Can G\"ormez and Dieter Van den
Bleeken on related topics leading up to
\cite{Figueroa-OFarrill:2022tlf}. In addition, we are grateful to an
anonymous referee for a critical reading of the manuscript and a
number of insightful comments and suggestions.

SP was supported by the Leverhulme Trust Research Project Grant
(RPG-2019-218) ``What is Non-Relativistic Quantum Gravity and is it
Holographic?''.

\pdfbookmark[1]{References}{ref}
\LastPageEnding

\end{document}